\documentclass[referee]{aa}
\usepackage{graphics}

\begin{document}

\thesaurus{10(10.01.1;10.05.1;10.06.1;10.06.2)} 

\title{Two-component model for the chemical evolution of the Galactic disk}

\author{R.X. Chang, J.L. Hou, C.G. Shu and C.Q. Fu}

\offprints{R.X. Chang,  E-mail: crx@center.shao.ac.cn }

\institute{Shanghai Astronomical Observatory, Chinese Academy of Sciences, Shanghai, 200030, China} 

\date{}

\titlerunning{Two-component model of Galactic chemical evolution}
\authorrunning{R.X. Chang et al}
   
\maketitle

\begin{abstract}
In the present paper, we introduce a two-component model of the Galactic 
disk to investigate its chemical evolution. The formation of the 
thick and thin disks occur in two main accretion episodes with both 
infall rates to be Gaussian. Both the pre-thin and post-thin scenarios for 
the formation of the Galactic disk are considered. The best-fitting is 
obtained through $\chi^2$-test between the models and the new observed metallicity 
distribution function of G dwarfs in the solar neighbourhood (Hou et al 1998). 
Our results show that post-thin disk scenario for the formation of the Galactic disk 
should be preferred. Still, other comparison between model predictions and 
observations are given. \\

\end{abstract}

\section{Introduction}

Since the existence of the thick disk of our Galaxy was confirmed by 
Gilmore \& Reid (1983) more than ten years ago, it has been generally 
accepted that a complete description of the thick disk, such 
as its scale length, scale height, density normalization, metallicity and 
kinematical properties, is a necessary step towards understanding the 
Galaxy formation, halo collapse, disk dynamical and chemical evolution. 
Unfortunately, the characteristics of this population remain controversial, 
especially for its density profile. Several attempts have been made to
deduce this parameter by remote star counts and field star survey ( Buser
\& Kaeser 1985, Gilmore, Reid \& Hewett 1985, Reid \& Majewski 1993, Buser,
Rong \& Karaali 1998). But the results are still quite uncertain. This is partly 
due to the lack of complete sample of thick disk stars since its members 
cannot be easily recognized from that of the thin disk and/or the halo 
in most observable distributions. Moreover, the determination of thick disk 
characteristics requires large star samples in various directions well 
distributed in both the longitude and latitude (Robin et al. 1996), which 
cannot be obtained easily at the present time. 

Recently, the study of chemical evolution of the Galactic disk has been proven
to be a powerful tool to explore the formation and evolution of our Galaxy.
Numerous models have been detailedly put forward (Matteucci \& Francois 1989, 
Ferrini et al. 1994, Giovagnoli \& Tosi 1995, Prantzos \& Aubert 1995, 
Timmes et al. 1995, Carig 1996, Pilyugin \& Edmunds 1996a,
Chiappini, Matteucci \& Gratton 1997, Allen et al. 1998, 
Thon \& Meusinger 1998, Prantzos \& Silk 1998).
Among them, Chiappini, Matteucci \& Gratton (1997 hereafter CMG97) were the 
first to take into account the effect of thick
disk. They assumed that there are two main accretion episodes. The first
is responsible for the formation of the thick disk, and the second, delayed
relative to the first, forms the thin disk. The predictions of their best-fitting model 
are in good agreement not only with the observed metallicity distribution 
, but with the observed number of very low metallicity stars (Rocha-Pinto 
\& Maciel 1996, hereafter RM96). This enlightens us to do more detailed analyze 
of the disk evolution based on the new chemical constraints.

In the present paper, the two-component model for the Galactic disk evolution 
(such as CMG97) is adopted, in which the local surface density of 
the thick disk at the present time is chose to be one of the 
free parameters. The infall rate is assumed to be in a Gaussian form instead of 
an exponentially decreasing one. The quantitative comparison between model 
predictions and the observations, i.e., the new G-dwarf metallicity distribution 
obtained by Hou et al (1998), is used for the $\chi^2$-test of the best-fitting model.

The outlines are as follows. In section 2, we present brief 
description of observational constraints up to now, of which the most 
important is the G-dwarf metallicity distribution. Section 3 is the model 
and its main ingredients. In section 4, we present best-fittings of four 
different models, which are closed-box, one-component, pre-thin and post-thin 
models respectively, to the observations. Discussions of the models are also included in 
section 4. Our conclusions are shown in the last section.  \\

\section {Observational constraints}

A successful model of the chemical evolution of the galactic disk should
reproduce the main observational features of both the solar neighborhood
and whole disk. Our set of constraints includes: \\
(1) G-dwarf metallicity distribution in the solar neighbourhood (Hou et al 1998); \\
(2) radial abundance gradients at present time; \\
(3) age-metallicity relation (AMR); \\
(4) the correlation between [O/Fe] and [Fe/H]; \\
(5) radial profiles for the gas surface density; and \\ 
(6) the variations of Star Formation Rate (SFR) across the disk. \\

The first one is selected as the observational constraint to quantitatively 
estimate the best-fitting model in this paper, since G dwarfs cover 
the whole life of the Galactic disk and its metallicity distribution can reflect 
the local chemical enrichment history. The others are used for the comparisons 
between the best-fit model predictions and observations.
 
\subsection {G-dwarf metallicity distribution} 

The metallicity distribution of G dwarfs in the solar neighborhood is one
of the most important constraints on the chemical evolution of the Galactic
disk. Since G dwarfs have lifetimes comparable to the estimated age of
the Galaxy, they represent a sample which has never been depleted by
stellar evolution, accumulating since the first episodes of low-mass star
formation. Therefore, a complete sample of these stars in the solar
neighborhood carries memory of the local star formation and chemical
enrichment history. 

Pagel \& Patchett (1975) derived a cumulative G-dwarf metallicity
distribution, based on a volume-limited sample of $132$ G dwarfs within about
 25 pc of the Sun. Pagel (1989) revised previous data of Pagel
\& Patchett (1975) by means of a new calibration between the ultraviolet
excess $\delta$ (U-B) and [Fe/H]. Later, Rana (1991) and Sommer-Larsen (1991)
independently revised the distribution of Pagel (1989), taking into account
the dynamical heating effect on the observed distribution. 
RM96 derived a G-dwarf metallicity distribution in the solar neighborhood, 
using $uvby$ photometry and up-to-date parallaxes. RM96 introduced a chemical 
criterion, according to which all sters have [Fe/H] $<$ -1.2 are considered to be 
halo members and excluded from the final sample.
The distribution of RM96 comprises 287 G dwarfs 
within 25 pc from the Sun and differs from the classic one by having a 
prominent single peak around [Fe/H] = - 0.2 (see RM96 for details).  

Recently, Hou et al (1998) collected a new, enlarged sample of G dwarfs within 
25pc from the Sun. The stars are selected from the third Catalogue 
of Nearby Stars (Gliese \& Jahreiss 1991). The $uvby$ data are taken from 
the catalogues of Olsen (1993) and Hauck \& Mermilliod (1990). No chemical criterion 
was introduced in Hou et al (1998) since observational evidences showed that, in the 
metallicity interval -1.5 $<$ [Fe/H] $<$ -1.0, the fraction of thick disk stars in the 
solar neighbourhood appears to be as high as 60$\%$ (Nissen \& Schuster 1997). This is 
one of the main differences between the distribution of Hou et al (1998) and that 
of RM96. The final sample contains 382 G dwarfs 
with photometric data. The adopted metallicity calibration and kinetic 
correction is the same as that of RM96. Following 
Pagel (1989) and RM96, Hou et al (1998) have also 
corrected the distribution for observational errors and cosmic scatter. 
\begin{table}
\caption{The obtained G-dwarf metallicity distribution in the solar neighbourhood 
taken from Hou et al (1998)} 
\begin{center}
\noindent
\begin{tabular}{lcccc}
\hline \hline
[Fe/H] & $\Delta N_{0}$ & f & $\delta (\Delta N_{0})$ & 
$\frac{\Delta N_{0}/{f}+ \delta (\Delta N_{0})}{N}$  \\ 
 \hline                               
 -1.5 $\sim$ -1.4 & 1 & 0.23 & -0.163 & 0.0091508 \\
 -1.4 $\sim$ -1.3 & 2 & 0.23 & -1.276 & 0.0110225  \\
 -1.3 $\sim$ -1.2 & 2 & 0.23 & -0.422 & 0.0110225 \\
 -1.2 $\sim$ -1.1 & 0 & 0.23 & -0.571 & 0.0110225 \\
 -1.1 $\sim$ -1.0 & 1 & 0.23 & -0.670 & 0.0177367 \\
 -1.0 $\sim$ -0.9 & 0 & 0.23 & -0.646 & 0.0177367 \\
 -0.9 $\sim$ -0.8 & 5 & 0.23 & -0.437 & 0.0177367 \\
 -0.8 $\sim$ -0.7 & 8 & 0.36 & -0.030 & 0.0485269 \\
 -0.7 $\sim$ -0.6 & 12 & 0.53 & 0.513 & 0.0506311 \\
 -0.6 $\sim$ -0.5 & 22 & 0.79 & 1.055 & 0.0632013 \\
 -0.5 $\sim$ -0.4 & 22 & 0.85 & 1.430 &  0.0597229 \\
 -0.4 $\sim$ -0.3 & 44 & 0.98 & 1.511 & 0.1014807   \\
 -0.3 $\sim$ -0.2 & 56 & 0.99 & 1.270 & 0.1264671 \\
 -0.2 $\sim$ -0.1 & 62 & 1.00 & 0.790 & 0.1373006  \\
 -0.1 $\sim$ 0.0 & 58 & 1.00 & 0.217 & 0.1273009 \\
 0.0 $\sim$ 0.1 & 47 & 1.00 & -0.260 & 0.1022046  \\
 0.1 $\sim$ 0.2 & 29 & 1.00 & -0.569 & 0.0621690  \\
 0.2 $\sim$ 0.3 & 10 & 1.00 & -0.678 & 0.0203841  \\ 
 0.3 $\sim$ 0.4 & 3 & 1.00 & -0.630 & 0.0051824 \\
\hline \hline
\end{tabular}
\end{center}
\end{table}
The results of Hou et al (1998) are shown in Table 1, in which the sample is divided into 
19 bins. The first column of Table 1 
is the metallicity range in each bin. The 
raw distribution $(\Delta N_0)_i$ are presented in column 2. 
The third is the weight factor $f_i$ for the scale height correction 
according to Sommer-Larson (1991). The fourth presences 
$\delta(\Delta N_{0})_i$, the correction 
factors for the observational errors and cosmic scatter. The obtained 
relative distribution is given in the last column, where 
$N = \sum_{i=1}^{19} [\frac{(\Delta N_0)_i}{f_i}+\delta(\Delta N_0)_i]$ 
is the total number of G dwarfs after correction. In the last column, 
the second to fourth bins are grouped to yield a mean value for the 
distribution, which is similar with the method used in RM96. This procedure 
is also used for the fifth to seventh bins. Table 
1 shows that the resulted distribution of Hou et al (1998) differs from that 
of RM96 by having a larger width and smaller amplitude of the single peak. 
Moreover, the metal-poor tail of the new distribution (Hou et al 1998) extends 
to [Fe/H]=-1.5.

\subsection { Abundance gradients}

Furthermore, radial metallicity variations of the interstellar medium 
(ISM) can constrain models of Galaxy formation and chemical evolution. 
From extensive studies of optical emission lines in HII regions, Shaver
et al. (1983) derived an oxygen abundance gradient of the order of
 -0.07 dex/kpc. Afflerbach et al.(1996) have indirectly deduced a
similar result by measuring electron temperature
variations in a set of ultra-compact HII regions. A relatively flatter
gradient has been obtained by Vilchez \& Esteban(1996) for the outer
Galaxy, using spectroscopic observations of a sample of HII regions
towards the Galactic anti-center. 

The recent radial profile of oxygen in the Galaxy can also be traced 
by observations of B-type stars, with main sequence ages less then 
1.0 Gyr. A series of medium- to high-resolution spectroscopic 
observations of early B-type main-sequence objects have been published 
(Smartt et al. 1997 and references there in). Using this homogeneous 
sample, Smartt et al.(1997) derived an oxygen abundance gradient of 
-0.07 $\pm$ 0.01 dex/kpc between Galactocentric distance of 
6 kpc $\le r \le$ 18 kpc, which is in 
good agreement with nebular studies. Gummersbach et al (1998) 
determines the stellar parameters and abundances of several element for 
16 early B main-sequence stars in Galactocentric distance 5.0 kpc $\le 
r \le$ 14.0 kpc by reanalyzing and extending the observations of Kaufer 
et al (1994). An oxygen abundance gradient  -0.07 $\pm$ 0.01 dex/kpc  
is derived, typical for normal spiral galaxies of similar Hubble type. 

\subsection {Others}

Just as mentioned above, other chemical constraints should be taken into 
account at the same time, such as the age-metallicity relation, the 
correlation between [O/Fe] and [Fe/H] for field stars as well as the radial profiles 
of gas surface density and SFR at the present time for the whole disk.

Twarog (1980) obtained the first AMR for the local disk stars, with the
stellar ages from the theoretical isochrones. The same sample has been
reanalyzed by Carlberg et al. (1985) with different results, due to
the revision of the isochrones as well as the calibration of abundances.
A more accurate AMR was obtained by Edvardsson et al. (1993). They have 
derived abundances of 13 different elements, such as O, Fe, Si, Ba etc., 
as well as individual photometric ages, for $189$ nearby
field F and G dwarfs. Their abundance analysis was made with theoretical 
LTE model atmospheres, based on the extensive high resolution, high S/N, 
spectroscopic observations of carefully selected field stars. The resulted 
AMR of Edvardsson et al. (1993) was used at the present study. However, 
this AMR does not constitute a tight constraints of the chemical model, 
since there is a considerable scatter. Moreover, the results of the survey 
of Edvardsson et al. (1993) concerning O vs. Fe relationships for field 
stars are used in this study. As for metal-poor stars, the 
correlation between [O/Fe] and [Fe/H] are taken from Barbuy (1988).

The radial Galactic profiles of atomic and molecular hydrogen are 
discussed in Lacey \& Fall (1985). An updated discussion is given in 
Dame (1993). Inside the solar circle, the molecular and atomic gas are 
found in roughly equal amounts. However, the surface density of atomic 
hydrogen, which seems to be constant from $4kpc$ to $15kpc$, dominates the 
gas profiles outside the solar circle. The radial distribution of the 
sum of atomic and molecular hydrogen given in Dame (1993) is adopted in 
this paper. 
 
The radial distribution of the present SFR in the Galaxy are taken from 
Gusten \& Merger (1983), Lyne et al (1983) and Guibert et al (1978). 
Data are based on several tracers of star formation: Lyman continuum 
photons from HII regions, pulsars and supernova remnants. It is normalized 
to the present SFR in the solar neighbourhood (as in Lacey \& Fall 1985) 
since the absolute values depend on poorly known conversion factors.

\section {The model }

It is assumed that the Galactic disk is sheet-like, which 
originates and grows only from the infall of primordial gas. The disk
is considered as a system of independent rings with $1kpc$ wide for each. 
No radial inflows or outflows are considered and the center of each ring 
locates at its median Galactocentric radius. The ring centered at Galactocentric
distance $r_\odot$ = 8.5 kpc is labeled as the solar neighbourhood. The 
age of the disk is adopted to be 13.0Gyr (Rana 1991).

\subsection{Basic equations and nucleosynthesis} 

The instantaneous-recycling approximation (IRA) is relaxed, but 
instantaneous mixing of the gas with the stellar ejecta is assumed, i.e., 
the gas is characterized by a unique composition at each epoch of time. 
We solve numerically the classical set of equations of Galactic chemical 
evolution (Tinsley 1980, Pagel 1997) as
\begin{eqnarray}
\frac{d\Sigma_{tot}(r,t)}{dt} & = & f(r,t), \\ 
\frac{d\Sigma_{gas}(r,t)}{dt} & = & -\psi(r,t)+ \int_{m_t}^{100}(m-m_{r})
\psi(r,t-\tau_m)\phi(m)\, {\rm d}m +f(r,t), \\
\frac{d[Z_{i}(r,t)\Sigma_{gas}(r,t)]}{dt} & = & -Z_{i}(r,t)\psi(r,t)+ 
\int_{m_t}^{100}my_{i,m}
\psi(r,t-\tau_m)\phi(m)\, {\rm d}m +Z_{i,f}f(r,t),
\end{eqnarray}
where $\Sigma_{tot}(r,t)$ and $\Sigma_{gas}(r,t)$ are the total and gas 
surface density respectively in the ring centered at Galactocentric distance $r$ at 
evolution time $t$; $f(r,t)$ is often called the infall or accretion rate; 
$\psi(r,t)$ is the star formation rate (SFR) and $\phi(m)$ is the initial mass 
function (IMF); $m_r$ and $\tau_m$ are the remnant mass and the lifetime of 
a star of initial mass $m$, respectively, and $m_t$ is the corresponding initial 
mass of a star whose main-sequence lifetime $\tau_{m}$ equates to evolution time $t$ 
(the turnoff mass). Here the mass range of IMF is taken from $0.1 M_\odot$ to 
$100 M_\odot$. The mass of element $i$ in the gas evolves via star 
formation (putting metals from the ISM into stars), ejection, and gas inflows, 
according to equation (3), where $y_{i,m}$ is the stellar yield of element $i$, i.e., 
the mass fraction of a star of initial mass $m$ that is converted to element 
$i$ and ejected, and $Z_{i,f}$ is the mass abundance of element $i$ in the 
infalling gas, which is assumed to be piromordial in this study: $Z_{O,f}=
Z_{Fe,f}=0$. It should be emphasized that the second terms in the right 
hand of equations (2) and (3) also include the contribution of Type Ia 
supernovas (Type Ia SNs), which is detailedly presented in Matteucci \& 
Greggio (1986). The constant $A$ in equation (9) of Matteucci \& 
Greggio (1986) describes the fraction of systems with total mass in appropriate 
range, which eventually succeed in giving rise to a Type Ia SN event, and in this study, 
it is fixed by requiring to present best-fit to the metal-rich tail of the G-dwarf 
metallicity distribution in the solar neighbourhood.  

It is also assumed that every star ejects its envelope just after leaving 
the main sequence. The adopted relation between main-sequence lifetimes $\tau_m$
( in units of Gyr) and stellar initial mass $m$ (in units of $M_\odot$) is ( Larson 1974):
\begin{equation}
log m = 1.983 - 1.054 \sqrt {(log\tau_{m} +2.52)}.
\end{equation} 
For the sake of simplicity, we assume that, except for Type Ia SNs, any star
evolves as a single star even if it is the member of a binary system. All
massive stars ($m > 9 M_\odot$) explode as type II supernovas (Type II SNs), 
leaving behind a neutron star of mass $m_R=0.5 M_\odot$ (Prantzos \& Silk 1998). 
The final stage of the intermediate /low mass stars ($M \leq 9 M_\odot$) is
white dwarfs, and the final-initial mass relation is taken from 
Weidemann(1984). Type Ia SNs are thought to originate 
from carbon deflagration in C-O white dwarfs in binary systems. The method 
included the contribution of Type Ia SNs is the same as 
Matteucci \& Greggio (1986). 

In this paper, we only consider the evolution of iron and oxygen. The
oxygen and iron production for Type II SNs and Type Ia SNs are taken from Woosley 
\& Weaver (1995) and Woosley (1997), respectively. Recently, using the
evolutionary tracks of Geneva group up to the early asymptotic giant
branch (AGB) in combination with a synthetic thermal-pulsing AGB model,
van den Hock \& Groenewegen (1997) calculated in detail the chemical
evolution and yields of six elements up to the end of AGB. Their results
showed that the low-mass stars ($m < 3M_\odot $) produce small amounts
of oxygen, yet the intermediate mass stars ($3M_\odot \leq m \leq
8M_\odot$) destroy the initial oxygen through Hot Bottom Burning (HBB).
Therefore, it is reasonable in this paper to neglect the oxygen production by
intermediate/low mass stars compared with that of massive stars.  

\subsection{The infall rate}

Currently popular models of the galaxy formation are semi-analytic 
models within the framework of the hierarchical structure formation 
paradigm (White \& Rees 1978, White \& Frenck 1991, Wechsler et al 
1998, Mo et al 1998), which allow one to model the astrophysical 
processes involved in galaxy formation in a simplified but physical 
way (Kauffmann et al 1993, Somerville \& Primack 1998, Kauffmann et 
al 1998, Primack et al 1998). These models are in good agreement with 
a broad range of local galaxy observations, including the correlation 
between luminosity and circular velocity for spirals ( the Tully-Fisher relation), 
the B-band luminosity function, cold gas contents, metallicities, 
and colors. These models postulate that the formation of galaxy is 
mainly regulated by gas cooling, dissipation, star formation, and 
supernova feedback. However, for the purpose of simplicity, 
Galactic chemical evolution models assume that the integral effect of 
these processes can be represented by that of an infall rate, which 
is a function of evolution time and Galactocentric distance.
The form of the infall rate is adjusted to 
satisfy the constraint of G-dwarf metallicity distribution in the solar 
neighbourhood. Although an exponentially decreasing infall rate is 
widely used, we adopt here a Gaussian form for the infall rate. The 
physical motivation for such a choice is that because of its small initial 
surface density, the local disk initially accretes a small amount of the 
surrounding gas; as the disk mass and gravitational potential build up, 
the accretion rate gradually increases, but starts decreasing when the 
gas reservoir is depleted (Prantzos \& Silk 1998).

Numerous models for the formation of thick disk have been put forward since 
the confirmation of its existence by Gilmore \& Reid (1983) (see Majewski 1993 
for details). The models fall into either "top-down" scenarios (the pre-thin 
model), where the formation of the thick disk precedes that of the thin disk, 
or the "bottom-up" scenarios (the post-thin model), where the thick disk is 
the result of some action on or by 
the thin disk. In the pre-thin disk model, the formation of the thick disk 
is a transitional phase during the general contraction of the Galaxy. This 
model views the thick disk as a dissipative, rotational-supported structure, 
and the halo as non-dissipative and supported by the kinetic pressure provided 
by large, anisotropic velocity dispersions. The post-thin model resorts to 
formation of the thick disk after the gas has completely collapsed into a thin 
disk. Possible physical processes are: (1) Secular kinematic diffusion of thin 
disk stars (Norris 1987); (2) Violent thin disk heating by the accretion of 
a satellite galaxy (Quinn et al 1993). The required events must not occur too 
late in the disk life time so that the gas can cool again and form stars in the 
thin disk; (3) Halo response to disk potential (Gilmore \& Reid 1983). The post-thin 
model leaves two important observational signatures. First, the thick disk is a 
separate population distinct from the thin disk and the halo. Second, no gradient 
can be generated in the thick disk by the events, although a pre-existing gradient 
may survive the merger.

In this study, both the pre-thin and post-thin model for the formation of 
Galactic disk are considered. Following CMG97, we assume that 
there are two main infall episodes in both cases.
The rate of mass accretion (in unit of 
$M_{\odot} pc^{-2}Gyr^{-1}$) in each ring could be expressed as 
\begin{equation}
f(r,t)=\frac{A(r)}
{\sqrt{2\pi}\sigma_t}
e^{-(t-\tau_t)^{2}/{2\sigma_t^2}} + \frac{B(r)}{\sqrt{2\pi}\sigma_d}
e^{-(t-\tau_d)^{2}/{2\sigma_d^2}},
\end{equation}
where $\tau_t$ and $\tau_d$ (in units of Gyr) are the maximum infall time of 
the thick and thin disk respectively, $\sigma_t$ and $\sigma_d$ (in units of Gyr) 
are the corresponding half-width. It is assumed that $\sigma_t \approx 
\tau_t$ and $\sigma_d \approx \tau_d$, i.e., the value of the 
half-width can just varies a little 
around that of the formation time-scale (Prantzos \& Silk 1998). 

For the pre-thin model, the first infall episode forms the thick disk, 
which originated from a fast dissipative
collapse such as that suggested by Eggen, Lynden-Bell \& Sandage (1962)
(Sandage 1990, Majewski 1993, CMG97). The second infall episode, delayed
to the first, forms the thin disk component, with a time-scale much longer
than that of the thick disk. The infall rate $f(r,t)$ is normalized to 
the local disk density at present time, i.e., \(\int_{0}^{t_g} 
f(r,t)\, {\rm d}t = \Sigma_{tot}(r,t_g) \). Assuming 
the total masses of different rings in the disk at the present time are all 
exponentially decreased with the increasement of Galactocentric distance 
for thin and thick disk with the same scale-length $r_0$, 
the form of $A(r)$ and $B(r)$ in the pre-thin model can be written respectively as
\begin{eqnarray}
A(r) & = & \frac{\Sigma_{thick}(r_\odot, t_g)}
{\int_{0}^{t_g} \frac{e^{-(t-\tau_t)^2/{2\sigma_t^2}}}{\sqrt{2\pi}\sigma_t}
 \, {\rm d}t } e^{-\frac{r-r_\odot}{r_0}}  \\ 
B(r) & =  & \left \{ \begin{array}{ll} 0 & {\rm if\ } t < t_{max} \\
\frac{\Sigma_{tot}(r_\odot, t_g)-\Sigma_{thick}(r_\odot, t_g)}
{\int_{t_{max}}^{t_g} \frac{e^{-(t-\tau_d)^2/{2\sigma_d^2}}}{\sqrt{2\pi}\sigma_d}
 \, {\rm d}t } e^{-\frac{r-r_\odot}{r_0}} & {\rm if\ } t_{max}\le t\le t_g 
\end{array}
\right.    
\end{eqnarray}
respectively, where $t_g$ is the age of the Galactic disk, $t_{max}$ represents 
the epoch of time at which the formation of the thin disk 
begins, $\Sigma_{tot}(r_\odot, t_g)$ is the present total surface density in 
the solar neighborhood, and $\Sigma_{thick}(r_\odot, t_g)$ is the local surface 
density of the thick disk at the present time, which is one of the 
free parameters in our model. We adopted $r_0$ = 2.7 kpc (Robin et al 
1996, Kent 1992), and $\Sigma_{tot}(r_\odot, t_g)$ = 55.0 
$M_\odot pc^{-2}$ (Rana 1991, Sackett 1997) for Galactic disk. There are four free parameters 
in the pre-thin model, $\tau_t, \tau_d, t_{max}$ and $\Sigma_{thick}(r_\odot, t_g)$. 

Contrary to the pre-thin model, the post-thin model assumes the first infall 
episode forms thin disk. Then, the thick disk forms delayedly as a result of some 
actions on or by the thin disk. Using the same method described above, the 
form of $A(r)$ and $B(r)$ in the post-thin model can be written as
\begin{eqnarray}
A(r) & =  & \left \{ \begin{array}{ll} 0 & {\rm if\ } t < t_{max} \\
\frac{\Sigma_{thick}(r_\odot, t_g)}
{\int_{t_{max}}^{t_g} \frac{e^{-(t-\tau_d)^2/{2\sigma_d^2}}}{\sqrt{2\pi}\sigma_d}
 \, {\rm d}t } e^{-\frac{r-r_\odot}{r_0}} & {\rm if\ } t_{max}\le t\le t_g, 
\end{array}
\right. \\
B(r) & = & \frac{\Sigma_{tot}(r_\odot, t_g)-\Sigma_{thick}(r_\odot, t_g)}
{\int_{0}^{t_g} \frac{e^{-(t-\tau_t)^2/{2\sigma_t^2}}}{\sqrt{2\pi}\sigma_t}
 \, {\rm d}t } e^{-\frac{r-r_\odot}{r_0}},     
\end{eqnarray}
respectively, where each parameter has the same notation as Eq. (6) and (7) 
except $t_{max}$, which represents the epoch of time at which the thick disk 
begins to form. There are also four free parameters 
in the pre-thin model, $\tau_t, \tau_d, t_{max}$ and $\Sigma_{thick}(r_\odot, t_g)$.

\subsection{SFR and IMF}

In the majority of chemical evolution models, star formation rate (SFR) is
assumed to depend on some power of the gas surface density ( Prantzos \&
Aubert 1995, Tosi 1996, CMG97). Based on gravitational instability, Wang
\& Silk (1994) developed a self-consistent model to derive the global star
formation rate as a function of radius in galactic disks. The resulted star
formation rate not only depends on the gas surface density, but also is
proportional to the epicycle frequency $\kappa$. Since $\kappa \propto
r^{-1}$, we adopt a similar star formation rate as that of Prantzos \&
Aubert (1995), which can be expressed as (in units of $M_\odot pc^{-2} Gyr^{-1}$ ) :
\begin{equation}
\psi(r, t) = \nu \Sigma_{gas}^{n}(r, t) /r 
\end{equation}
where $\Sigma_{gas}(r,t)$ and $r$ are in units of $M_\odot pc^{-2}$ and $kpc$, respectively.
The power law index $n=1.4$ is adopted (Prantzos \& Silk 1998), 
which is in some degree similar to that of Kennicutt (1998). The value of 
$\nu$ is derived from the condition of reproducing the present observed 
gas surface density in the solar neighborhood. We adopted 
$\Sigma_{gas}(r_\odot, t_g)$ = 10.0 $M_\odot pc^{-2}$ (Scoville \& Sanders 
1987, Prantzos \& Aubert 1995, Sackett 1997). 

The adopted stellar initial mass function (IMF) is taken from 
Kroupa et al. (1993), in which the IMF is described by a three-slope 
power law, $\phi(m) \propto m^{-(1+x)}$. In the high-mass region, 
the IMF has a relatively 
steep slope of $x$ = 1.7, while it flattens in the low-mass range ($x$ = 1.2 
for $0.5 M_\odot\le m \le 1.0 M_\odot$ and $x$=0.3 for $m < 0.5 M_\odot$). 
The adopted IMF is normalized to \(\int_{0.1}^{100.0} m\phi(m)\, {\rm d}m = 1 \).

\section {Results and discussions}

\subsection{$\chi^2$-test}

The observed G-dwarf metallicity distribution is treated as the 
tightest observational constraints on the chemical evolution models 
of the Galactic disk(CMG97). The metallicity of local disk stars 
in the sample of Hou et al (1998) extends to [Fe/H] = -1.5, 
which is consistent with the observations of the thick disk (Majewski 1993, 
Nissen \& Schuster 1997). However, the metal-weak tail of the thick disk is 
excluded from the distribution of RM96 by using a chemical criterion. Since 
the main aim of this work is to predict the general properties of 
the thick disk, we select the distribution of Hou et al (1998) as 
the observational constraint. 

The $\chi^2$ of the goodness-of-fit to the observed data is calculated as follows
\begin{equation}
\chi^2(n) = \sum_{i=1}^{n} (\frac{y_{mi}-y_{oi}}{\sigma_i})^2,
\end{equation}
where $y_{mi}$ and $y_{oi}$ are respectively the model-generated and 
observed data in $i$th data bin, $ \sigma_i $ is the error of the
observed data in $i$th bin, and  $n$ is the total number of bins. For
the metallicity distribution, Hou et al (1998) have already
divided the observed data into $19$ bins. The error $\sigma_i$ is given by
\begin{equation}
\sigma_i(\frac{\Delta N_i}{N}) = \sqrt{(\frac{\sqrt{\Delta N_i}}{N})^2 
                           + (\frac{\Delta N_i}{N} \frac{1}{\sqrt{N}})^2}
                           =\sqrt{\frac{\Delta N_i}{N}\frac{1}{N} 
                           + (\frac{\Delta N_i}{N} \frac{1}{\sqrt{N}})^2},
\end{equation}
where $\Delta N_i/N$ is the relative number of G-dwarfs in the $i$th bin
and $N$ is the total number of G-dwarfs in the corrected sample. 

\begin{table}
\caption{The results of best-fit to the new G-dwarf metallicity distribution 
for four different models}
\begin{center}
\noindent
\begin{tabular}{lccc}
\hline \hline
characteristics of & free parameters & $\chi^2$ & model confidential level   \\ 
   the model    &  (best-fit)  & (best-fit) & (best-fit)    \\
 \hline                               
 closed-box model:     &     &     &  \\
   $A(r)=0$ &    &   & \\
   $B(r)=0$  &  no  & 131.7  &  $< 0.1 \%$ \\ 
 one-component model:     &     &     &  \\
   $A(r)=0$  &    &   & \\
   $B(r)\neq 0$  & $\tau_{d}$= 3.8Gyr  & 20.3   & $30\%$  \\ 
 pre-thin model:     &     &     &  \\
   $A(r) \neq 0$  & $\tau_{t}$=1.0 Gyr   &   & \\
   $B(r)\neq 0$ (=0, if $t < t_{max}$) &  $\tau_d$=4.0Gyr  & 14.6  & $70\%$ \\
   $t_{max} \neq 0$ & $\Sigma_{thick}(r_{\odot},t_g)$=14.0$M_{\odot}pc^{-2}$  &   &  \\        
           & $t_{max}$=1.0Gyr &  & \\
 post-thin model:     &     &     &  \\
   $A(r) \neq 0$ (=0, if $t < t_{max}$)  & $\tau_{t}$=1.0 Gyr   &   & \\
   $B(r)\neq 0$ &  $\tau_d$=4.0Gyr  & 14.4  & $70\%$ \\
   $t_{max} \neq 0$ & $\Sigma_{thick}(r_{\odot},t_g)$=10.0$M_{\odot}pc^{-2}$  &   &  \\        
          & $t_{max}$=1.0Gyr &  & \\
\hline \hline
\end{tabular}
\end{center}
\end{table}

For comparison, we consider four different models respectively, 
which are closed-box, one-component, pre-thin and post-thin model. 
The differences among these models are the treatments of the infall 
rate. We perform model calculations for a broad range of free parameter combinations, 
if there are any, to present the best-fit to the new metallicity distribution. 
Using $\chi^2$-test, the results of best-fit to 
the G-dwarf metallicity distribution for these individual models are shown in 
table 2, which are detailedly discussed at following subsections. The 
characteristics of the models are shown in the first column. The 
resulted values of free parameters, the $\chi^2$ and the model confidential 
levels for the best-fit are shown in second, third, fourth column respectively. 
Figure 1 presents the best-fits to the new G-dwarf metallicity distribution 
for closed-box model (dotted line), the one-component model (dot-dashed line), 
pre-thin model (long dashed line) and post-thin model (full line). 

\subsection{Closed-box model}

The closed-box model considers the whole disk as an isolated system. Thus, 
$A(r)=B(r)=0$. No free parameter exists if the closed-box model is considered. 
Figure 2 shows that the closed-box model predicts larger number of metal-poor stars 
than that of the observations. It is often called the G-dwarf problem. The fact 
that closed-box model does not work is also shown in table 2, where the confidential 
level is too low to accept the model. Several possible explanations to the G-dwarf 
problem have been proposed (see brief reviews in Francois et al 1990, 
Malinie et al 1993). Generally, a good agreement between model predictions 
and observations is obtained by models that assume the disk formed by 
infall of piromordial gas (Pilyugin \& Edmunds 1996b, CMG97) 

\begin{figure}
\hspace{-10cm}
\resizebox{\hsize}{!}{\includegraphics{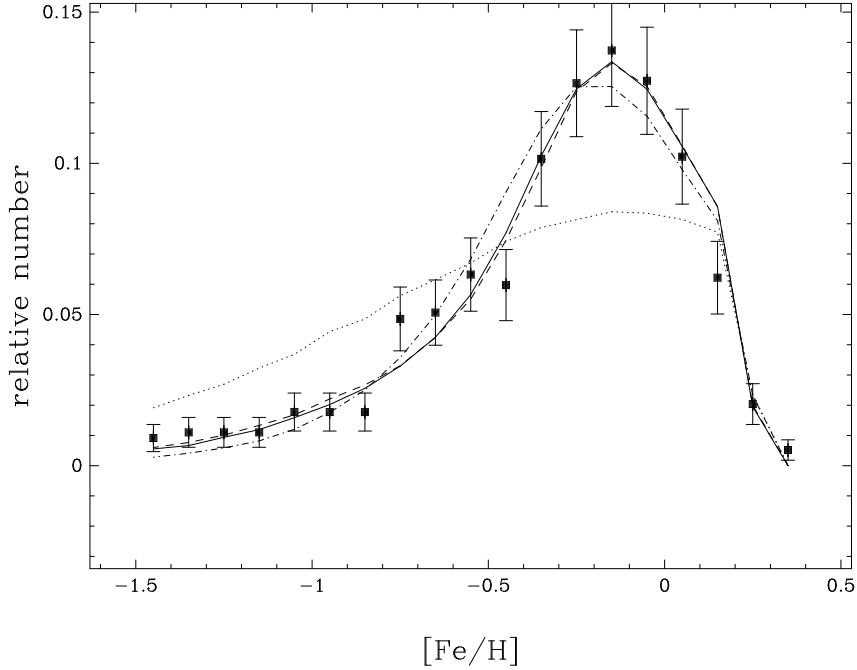}}
\caption{The best-fits to the new G-dwarf metallicity distribution 
for close-box model (dotted line), the one-component model (dot-dashed line ), 
pre-thin model (long dashed line) and post-thin model (full line). The observed 
data are taken from Hou et al (1998).}
\end{figure}

\subsection{One-component model}

The one-component model treats the thick and thin disk as one disk, which 
corresponds the infall rate of only one Gaussian form, i.e., $A(r)=0, B(r) \neq 0$. 
There is only one free parameter $\tau_d$, i.e., the infall timescale of the whole disk 
in one-component model after normalization mentioned above. Figure 1 shows that , for the 
one-component model, the goodness-of-fit between best-fit model predictions 
and observations seems to be acceptable, especially for the shape of the single 
peak of the metallicity distribution for G dwarfs. But, the results of $\chi^2$-test 
gives that the confidential level of one-component model is some 30$\%$, 
yet it can reach 70$\%$ for the two-component model (Table 2). This means one-component 
model is not the best, which is consistent with the observational result that the 
thick disk is kinematically and chemically different from the thin disk (see brief 
reviews in Majewski 1993). Therefore, it is necessary to treat the thick and thin 
disk differently, if one want to investigate detailedly the Galactic chemical evolution.

\subsection{Two-component model}

For the two-component model, both the pre-thin and post-thin model are 
considered. The differences between these two models are represented by different 
treatments of the infall rate, which are detailedly described in 
section 2.2. There are four free parameters in both the pre-thin and 
post-thin model, $\tau_t, \tau_d, t_{max}, \Sigma_{thick}(r_{\odot},
t_g)$. It should be emphasized that the $t_{max}$ has different meaning in 
different model. Figure 1 show that, for these two models, the best-fit model 
predictions are in good agreement with the observations, which are 
confirmed by the results of quantitative tests (Table 2). From 
Figure 1 and Table 2, it is difficult to distinguish which model is 
better. 

\begin{table}
\caption{Local density ratio of thick to thin disk from
literatures}
\begin{center}
\noindent
\begin{tabular}{lccc}
\hline \hline
Authors & scale height of thick disk & space density & $\Sigma_{thick}(r_{\odot},t_g)$   \\ 
     &  (pc)  &   normalization  &  $M_{\odot}pc^{-2}$\\
\hline                               
Gilmore \& Reid (1983)           &  1450             & 0.020    & 4.85    \\
Rose (1985)                      &  500 $\sim$ 1000  & 0.100   &  11.0 \\
Friel (1987)                     &  1000             & 0.050   &  7.85      \\
Sandage (1987)                   &  940              & 0.110    & 14.1   \\
Norris (1987)                    &  1100             & 0.034    & 6.09  \\
Kuijken \& Gilmore (1989)        &  1000             & 0.040    & 6.47   \\
Reid \& Majewski (1993)          &  1400 $\sim$ 1600 & 0.020 $\sim$ 0.025  &   5.67\\
Robin et al. (1996)              &  760              & 0.056 $\pm$ 0.010  & 6.83 \\
Ojha et al. (1996)               &  760              & 0.074 $\pm$ 0.020 & 8.69 \\
Buser et al. (1998)              &  1150             & 0.054 $\pm$ 0.015 & 9.43\\
\hline
This paper (1999):  &   &  &  \\
pre-thin model              &  1000           &   0.113  & 15.0  \\
post-thin model    & 1000 &  0.067  &  10.0  \\
\hline \hline
\end{tabular}
\end{center}
\end{table}

One important parameter predicted by the best-fit of our two-component 
model is the local surface density of thick disk at the present time. 
On the other hand, the density ratio of thick to thin disk are usually 
deduced from studies of star counts. In Table 3, we present the 
comparison between our model predictions and the data compiled from 
literatures. Since the density normalization and scale-height of the thick 
disk are anti-correlated when fitted simultaneously (Reid \& Majeweski 
1993, Majeweski 1993, Robin et al. 1996), the scale heights of the thick 
disk are also shown in 2th column. The previous results for the 
space density ratio of thick to thin disk (the parameter $D$) are shown 
in 3th column. Taking the scale-height of thick and thin disk as 
$h_t$ = 1000pc and $h_d$ = 300pc respectively as usual (Majewski 1993), 
the local space density ratio of thick to thin disk at the present time 
predicted by our model can be obtained based on the following equation:
\begin{equation}
D \approx \frac{\Sigma_{thick}(r_{\odot},t_g)}{\Sigma_{tot}(r_{\odot},t_g)
-\Sigma_{thick}(r_{\odot},t_g)} \frac{h_d}{h_t},
\end{equation}
where $\Sigma_{thick}(r_{\odot},t_g)$=10.0$M_\odot pc^{-2}$ and 15.0 
$M_\odot pc^{-2}$ for the best-fittings of the pre-thin and post-thin model, 
respectively. Moreover, equation (13) can also be used for deducing the $\Sigma_{thick}
(r_{\odot},t_g)$ with published $D$ and $h_t$ in literatures. The results 
are shown in 4th column of Table 3 with $h_d$=300pc and 
$\Sigma_{tot}(r_{\odot},t_g)$=55.0$M_\odot pc^{-2}$, respectively. 

\begin{figure}
\hspace{-10cm}
\resizebox{\hsize}{!}{\includegraphics{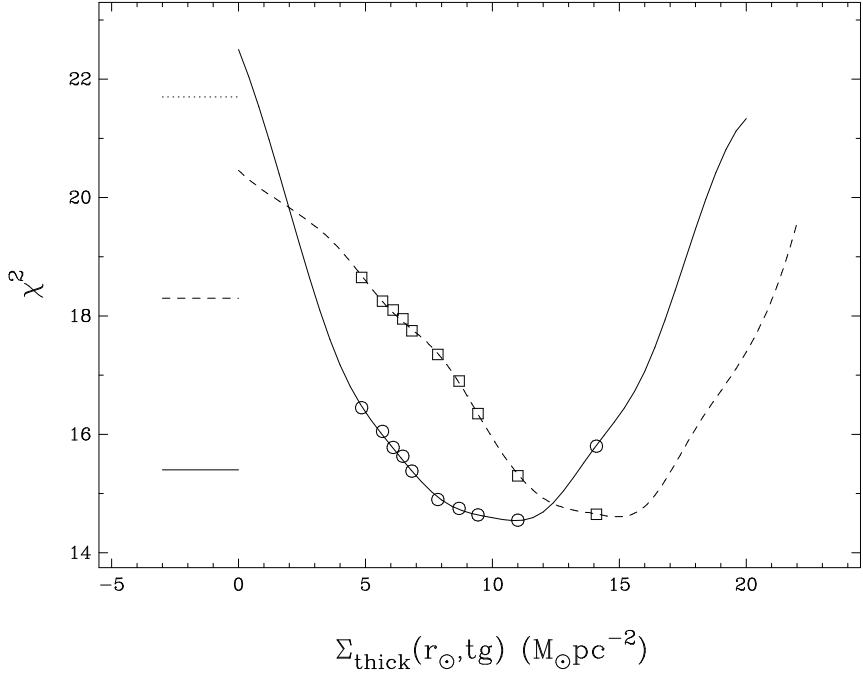}}
\caption{The $\chi^2$ as a function 
of the $\Sigma_{thick}(r_{\odot},t_g)$ within reasonable 
range for both the pre-thin model (long dashed curve) and post-thin model 
(full curve) in which other free parameters are fixed as $\tau_{t}$=1.0Gyr, 
$t_{max}$=1.0Gyr and $\tau_d$=4.0Gyr. Three short 
horizontal lines indicate the model confidential level of $30\%$ (dotted 
line), 50$\%$ (long dashed line) and 70$\%$ (full line), respectively. The points 
indicate the different results of $\chi^2$ if the value of 
$\Sigma_{thick}(r_{\odot},t_g)$ from ten literatures (Table 3) are adopted for both the pre-thin 
model (open squares) and post-thin model (open circles), respectively.}
\end{figure}

From Table 3, it can be seen that the previous density normalizations of thick disk 
span a wide range (from 0.02 to 0.11). This is partly due to
the difficulty in distinguishing the kinematical, chemical and spatial
characteristics of thick disk with halo and thin disk. Moreover, comparing
published star counts and color distribution in neighboring fields, star
density discrepancies are sometimes larger than photometric random errors
as established by authors (Ojha et al. 1996, Robin et al. 1996). But the results of 
more recent surveys are in good agreement within the error range (Ojha et al. 1996, 
Robin et al. 1996, Buser et al 1998).  Table 3 shows that the value of 
$\Sigma_{thick}(r_{\odot},t_g)$ predicted by the post-thin model are consistent 
with most of the previous results from literatures, while $\Sigma_{thick}(r_{\odot},t_g)$ 
predicted by the pre-thin model is larger than the previous data from studies of 
star counts. To illustrate this quantitatively, Figure 2 shows the $\chi^2$ as a function 
of the $\Sigma_{thick}(r_{\odot},t_g)$ within reasonable 
range for both the pre-thin model (long dashed curve) and post-thin model 
(full curve) in which other free parameters are fixed as $\tau_{t}$=1.0Gyr, 
$t_{max}$=1.0Gyr and $\tau_d$=4.0Gyr. It is shown that the value of $\chi^2$ 
is very sensitive to $\Sigma_{thick}(r_{\odot},t_g)$. This suggests that the thick disk 
has great influence on the Galactic chemical evolution. In figure 2, three short 
horizontal lines indicate the model confidential level of $30\%$ (dotted 
line), 50$\%$ (long dashed line) and 70$\%$ (full line), respectively. The points 
in Figure 2 indicate the different results of $\chi^2$ if the value of 
$\Sigma_{thick}(r_{\odot},t_g)$ from ten literatures (Table 3 ) are adopted for 
both the pre-thin 
model (open squares) and post-thin model (open circles), respectively.

Figure 2 shows that, for the post-thin model, five points have model 
confidential level larger than 70$\%$, while only two points have confidential 
level larger than 70$\%$ for the pre-thin model. This suggests that the post-thin 
model be better than pre-thin model. Other evidences tended to favour the post-thin 
scenario for the formation of thick disk comes from the following facts. First, 
the thick disk is kinematically distinct from the thin disk and it shows no 
kinematic gradients (Ojha et al 1994 a,b). Second, Gilmore et al (1995) studied 
metallicity distribution of thick disk stars up to about $3kpc$ from the 
Galactic plane. They found that thick disk stars show no vertical abundance gradient. 
This argues against dissipational setting as the formation process of the thick 
disk (Freeman 1996). 

\subsection{Discussions}

Based on the above discussions, we treat the post-thin model with $\tau_t$=1.0 Gyr, 
$t_{max}$=1.0Gyr, $\tau_d$=4.0 Gyr and $\Sigma_{thick}(r_{\odot},t_g)$=
10.0$M_\odot pc^{-2}$ as the best-fit model.

\begin{figure}
\hspace{-10cm}
\resizebox{\hsize}{!}{\includegraphics{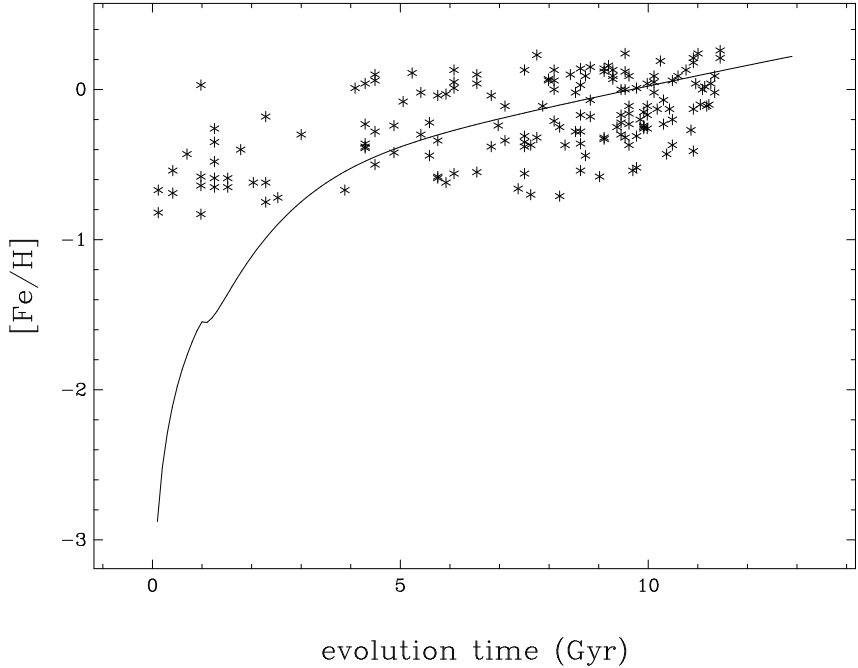}}
\caption{Temporal evolution of [Fe/H] predicted by the best-fit model 
(full line) for the solar vicinity. The observed data (asterisks) are from 
Edvardsson et al (1993). }
\end{figure}

Figure 3 presents the comparison between our best-fit model predictions for the 
AMR and the observations. The full line is our model predictions and the points 
are observational data taken from Edvardsson et al (1993). Figure 3 shows 
that, at the beginning of the formation of the thick disk ($t=t_{max}=1.0Gyr$), 
the iron abundance of ISM decreases a little due to the increasing infall 
rate of the primordial gas. After that phase, the model predicts that the 
metallicity increases smoothly with time. The overall tendency for this relation 
is consistent with the mean observations, but the present model can not 
reproduce the large observed scatters. Nordstrom et al. (1997) 
discussed in detail the main hypotheses for the origin of this scatter, such as 
star formation in an inhomogeneous gaseous medium, orbital diffusion in 
homogeneous galaxy and mergers or accretion events. However, a physical 
mechanism that reproduces the observed scatter in the AMR without violating 
other observational constraints has not yet been identified. 

\begin{figure}
\hspace{-10cm}
\resizebox{\hsize}{!}{\includegraphics{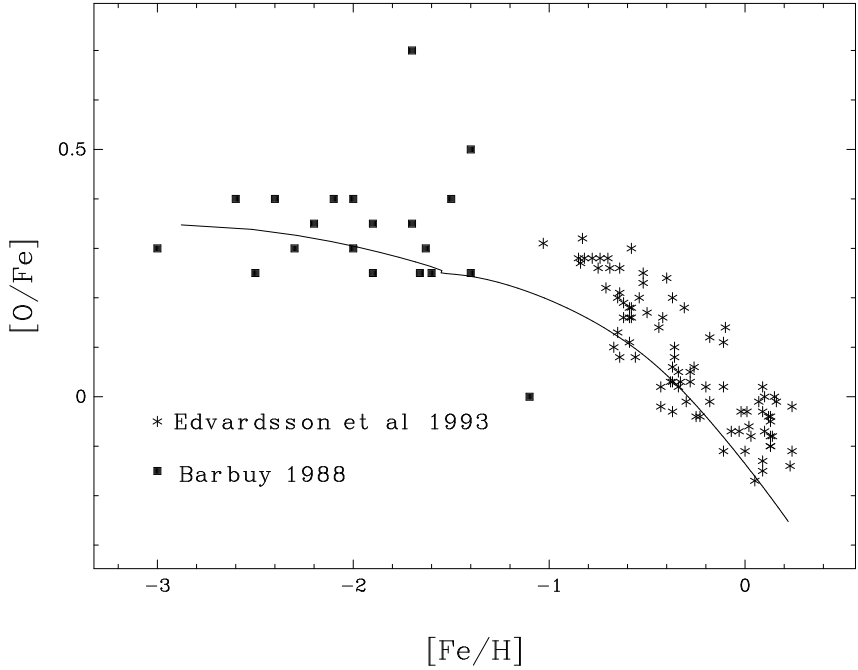}}
\caption{Predicted behavior of [O/Fe] vs. [Fe/H] for the best fit model (full 
line). The observed data are from Edvardsson et al (1993) (asterisks) and 
Barbuy (1988) (full squares).  }
\end{figure}
\begin{figure}
\hspace{-10cm}
\resizebox{\hsize}{!}{\includegraphics{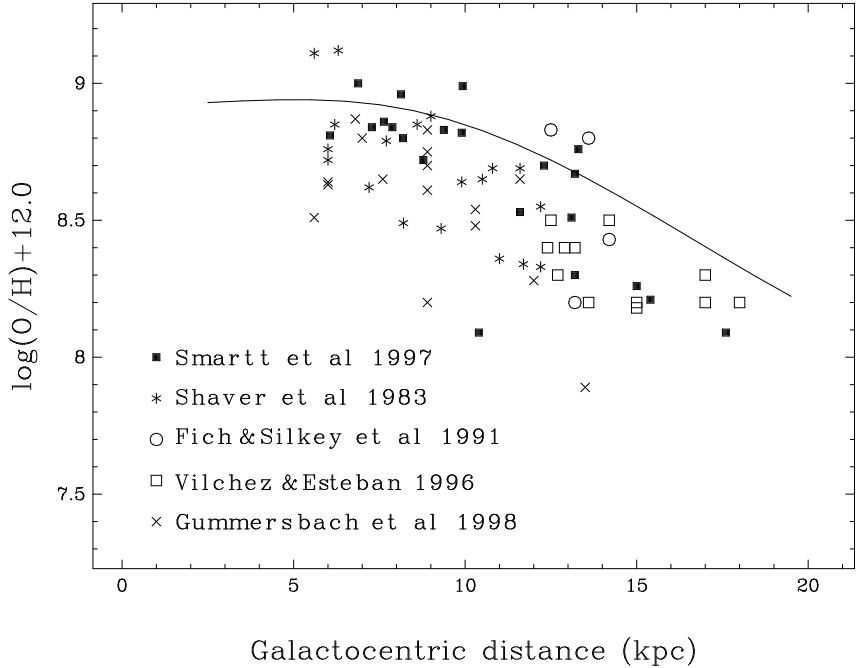}}
\caption{ Radial distribution of the oxygen abundance at the present time predicted 
by the best-fit model (full line). The observed data of HII regions are from Shaver
et al (1983) (asterisks), Fich \& Silkey (1991) (open circles) and Vilchez \& 
Esteban (1996) (open squares). The observations of early B-type main-squence stars 
 are taken from Smartt et al (1997) (full squares) and Gummersbach et al 
 (1998) (crosses). }
\end{figure}

Figure 4 compares the predicted behaviors of [O/Fe] vs. [Fe/H] for the best-fit 
model (full line) with the observations. The observed data are taken from 
Edvardsson et al (1993) (asterisks) and Barbuy (1988) (full squares). Our 
model predicts that there is a small loop at [Fe/H]= -1.6. The similar behavior 
is predicted in the best-fit model of CMG97. Figure 4 shows that our model 
prediction is in good agreement with the observations. This suggests that 
the relative stellar yield for oxygen and iron we adopted here be reasonable.

Contrary to the case of the solar neighbourhood, the available observations 
for the Milky Way disk offer information maily about its current status, 
not its past history. Therefore, there is much more freedom in constructing 
a model. Up to now, chemical evolution models of the Galactic disk consider 
the disk as a system of independent rings. This oversimplification generally 
ignores the possibility of radial inflows produced in gaseous disks, e.g. 
by viscosity or by the infall of gas with a specific angular momentum 
different from that of the underlying disk (Prantzos \& Silk 1998). 
Fortunately, a radial variation of the infall time-scale may play a similar 
role. In our best-fit model, the infall timescale of the thin disk $\tau_d$ 
is assumed to be radially dependent, taking an lower values in the inner disk 
($\tau_d$=2 Gyr at $r$=2 kpc) and larger ones in the outer disk ($\tau_d$=4 
Gyr at $r$=8.5 kpc). 

In Figure 5, we compare the predicted radial distribution 
of oxygen abundances by our best-fit model (full line) with the observations.
The observed data of HII regions are from Shaver et al (1983) (asterisks), 
Fich \& Silkey (1991) (open circles) and Vilchez \& Esteban (1996) (open 
squares). The observations of early B-type main-squence stars are taken 
from Smartt et al (1997) (full squares) and Gummersbach et al (1998) (crosses). 
It shows that model predictions of oxygen abundances are larger than 
most of the observations. Moreover, our model predicts that the abundance gradient 
in inner region (-0.01 dex/kpc for $r \le $ 8.5 kpc) is steeper than 
that in outer region (-0.07 dex/kpc for $r \geq $ 8.5 kpc). This 
seems to be contradictory to the observations 
of HII region, which suggests a flatter oxygen abundance gradients in 
outer region. This disagreement was also obtained by most of chemical 
evolution models for the Galactic disk (see brief reviews in Tosi 1996). 
This is probably due to the simplicities of the present models. Samland et 
al. (1997) developed a chemi-dynamical evolution model for the Galactic disk and 
presented better fit for the observed variations of oxygen abundance 
across the whole disk. Therefore, this disagreement  probably could be 
solved if one considers the influence of dynamical evolution on the 
chemical evolution and the effect of gas heating on star formation rate. 

Figure 6 presents the comparison between the predictions of our best-fit 
model for the radial profile of the present SFR (full line) and the 
observations. Observed data are normalized to SFR in the solar 
neighborhood. Data are based on several tracers of star formation: Lyman 
continuum photons from HII regions (full squares, from Gusten \& Merger 1983);
pulsars (open circles: from Lyne et al 1985); and supernova remnants 
(crosses: from Guibert et al 1978). The good agreement 
between our model prediction and the observations indicates that 
our model predicts reasonable star formation history for the Galactic disk.    

\begin{figure}
\hspace{-10cm}
\resizebox{\hsize}{!}{\includegraphics{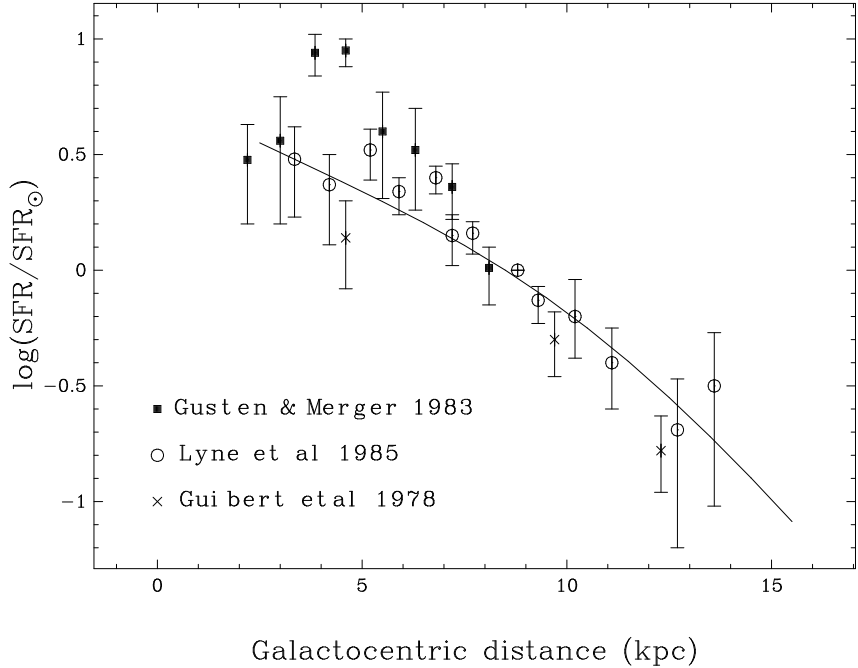}}
\caption{Profile of present star formation rate (SFR) in the galactic disk. Full 
line is the predictions of the best-fit model. Observed data are normalized to 
SFR in the solar neighborhood. Data are based on several tracers of star formation: 
Lyman continuum photons from HII regions (full squares, from Gusten \& Merger 1983);
pulsars (open circles: from Lyne et al 1985); and supernova remnants (crosses: from 
Guibert et al 1978). }
\end{figure}\begin{figure}
\hspace{-10cm}
\resizebox{\hsize}{!}{\includegraphics{fig7.ps}}
\caption{Radial distribution of the present gas surface density predicted by the 
best-fit model (full line). The lower dashed line 
is the sum of atomic and molecular hydrogen given in Dame (1993), corrected 
for the contribution of 30$\%$ helium. The upper one is obtained by adopting 
the gas surface density in the solar neighbourhood as 16 $M_{\odot}pc^2$ and 
scaling the curve of Dame (1993) accordingly (Prantzos \& Aubert 1995).}
\end{figure}

Finally, in Figure 7, we present the comparison for the radial 
distribution of the present gas surface density between our best-fit model 
predictions (full line) and the observations (dashed lines). Two dashed lines 
are reproduced from Prantzos \& Aubert (1995). The lower dashed line 
is the sum of atomic and molecular hydrogen given in Dame (1993), corrected 
for the contribution of 30$\%$ helium. The upper one is obtained by adopting 
the gas surface density in the solar neighbourhood as 16 $M_{\odot}pc^2$ and 
scaling the curve of Dame (1993) accordingly (Prantzos \& Aubert 1995). Given 
the uncertainties in the observational data, the model is also in good agreement 
with the observed profile.   

\section{Conclusions}

In this work, we introduce a two-component models for the chemical evolution 
of the Galactic disk, which assumes that the formation of the thick and thin 
disks occur in two main accretion episodes. The infall rate is assumed 
to be Gaussian. Both the pre-thin and post-thin scenarios for the formation 
of the Galactic disk are considered. The local surface density 
of the thick disk at the present time is chosen to be one of the free 
parameters. Following Prantzos \& Silk (1998), we also assume that the SFR 
is not only proportional to $n$ power of gas surface density, but 
directly correlates with the Galactocentric distance. Comparing model 
predictions with the new metallicity distribution in the solar neighbourhood 
(Hou et al 1998), we use the $\chi^2$-test to derive best-fittings and compare 
the reasonableness of four different models, which are closed-box, one-component, 
pre-thin and post-thin models. Moreover, comparisons between the predictions of 
our best-fit model and the main observational constraints are presented. 
The main results can be summarized as follows.

\begin{enumerate}
    \item Our results suggests that the post-thin model for the formation of 
    the Galactic disk should be preferred. This is consistent with the 
    observational evidences that the thick disk is chemically and kinematically 
    distinct from the thin disk and it shows no vertical abundance gradient. 

     \item The goodness-of-fit for model predictions about metallicity 
     distribution to the observations, $\chi^2$, 
     is very sensitive to the local surface density of the 
     thick disk at the present time. This suggests that it is necessary 
     to treat the thick and thin disks differently if one want to investigate 
     detailedly the Galactic chemical evolution. 
      
     \item The post-thin model predicts $\Sigma_{thick}(r_{\odot},t_g)$
     =10.0$M_\odot pc^{-2}$. The resulted space density ratio of thick 
     to thin disk is consistent with the previous data from recent 
     studies of star counts. However, the pre-thin model predicts a larger 
     value of the local thick disk density. 
     
     \item The predictions of our best-fit model are in good agreement not 
     only with the observed data in the solar neighborhood, but also with the main 
    observational features of the Galactic disk. However, contrary to 
    the observations in HII regions, our model predicts the 
    oxygen abundance gradient in outer region is steeper than that in 
    inner region. 

   \end{enumerate}

\begin{acknowledgements}
      The authors wish to thank Prof. Peng
Qiuhe of Nanjing University for helpful discussions. This research was
supported in part by a grant from the National Natural Science Foundation
of China and in part by the grant from Young Lab of Shanghai Observatory.
\end{acknowledgements}


\begin{thebibliography}{}

   \bibitem{} Afflerbach A., Churchwell E., Acord J.M. et. al. 1996, ApJS 106,423 
   
   \bibitem{} Allen C., Carigi L. \& Peimbert M., 1998, ApJ 494, 247 
   
   \bibitem{} Barbuy B., 1988, A\&A 191, 121 
   
   \bibitem{} Buser R. \& Kaeser U., 1985, A\&A 145, 1
    
   \bibitem{} Buser R., Rong J.X. \& Karaali S., 1998, A\&A 331, 934 
   
   \bibitem{} Carigi L., 1996, Rev.Mex.A.A. 32, 179 
   
   \bibitem{} Carlberg R.G., Dawson P.C., Hsu, T. \& VandenBerge D.A., 1985, ApJ 294, 674 
   
   \bibitem{} Chiappini C., Matteucci F. \& Gratton R., 1997, ApJ 477, 765 (CMG97) 
   
   \bibitem{} Dame T.M., in "Back to the Galaxy', eds. S.Holt and F. Verter (AIP), p.267
   
   \bibitem{} Edvardsson B., Andersen J., Gustafsson B. et al., 1993, A\&A 275, 101 
   
   \bibitem{} Eggen O.J., Lynden-Bell D. \& Sandage A.R., 1962, ApJ 136, 748 
   
   \bibitem{} Ferrini F., Molla M., Pardi C. \& Diaz A.I., 1994, ApJ 427, 745 
   
   \bibitem{} Fich M. \& Silkey M., 1991, ApJ 366, 107 
   
   \bibitem{} Francois P., Vangioni-Flam E. \& Audoze J., 1990, ApJ 361, 487
  
   \bibitem{} Freemann K.C., 1996, in IAU Symposium 171, New Light on Galaxy Evolution, 
   eds. R. Bender \& R.L. Davies, p.1
   
   \bibitem{} Friel E.D., 1987, AJ 93, 1388 
   
   \bibitem{} Gilmore G.F. \& Reid I.N., 1983, MNRAS 202, 1025 

   \bibitem{} Gilmore G.F., Reid I.N. \& Hewett P.C., 1985, MNRAS 213, 257 
   
   \bibitem{} Gilmore G.F., Wyse R. \& Jones B., 1995, AJ 109, 1095
   
   \bibitem{} Giovagnoli A. \& Tosi M., 1995, MNRAS 273, 499 
   
   \bibitem{} Guibert J., Lequeux J. \& Viallefond F., 1978, A\&A 68, 1

   \bibitem{} Gliese W., Jahreiss H., Third Catalogue of Nearby Stars. 
   Astron. Rechen-Inst. Heidelberg 
    
   \bibitem{} Gummersbach C.A., Kaufer A., Schafer D.R., Szeifert T. 
   \& Wolf B., 1998, A\&A 338, 881
    
   \bibitem{} Gusten R. \& Mezger M., 1983, Vistas Astr. 26, 159 
   
   \bibitem{} Hauck B. \& Mermilliod M., 1990, A\&AS 86, 107
   
   \bibitem{} Hou J.L., Chang R.X. \& Fu C.Q., 1998, Annals of Shanghai Observatory 
   Academia Sinica, No. 19, 96
   
   \bibitem{} Kauffmann G., White S.D.M. \& Guiderdoni B., 1993, MNRAS 264, 201
   
   \bibitem{} Kauffmann G., Colberg J.M., Diaferic A. \& White S.D.M., 1998, MNRAS in press (astro-ph 9805283)
      
   \bibitem{} Kennicutt R.C. Jr., 1989, ApJ 344, 685 
   
   \bibitem{} Kennicutt R., 1998, ApJ 498, 541
   
   \bibitem{} Kent S., 1992, ApJ 387, 181
   
   \bibitem{} Kroupa P., Tout C. \& Gilmore G., 1993, MNRAS 262, 545 
   
   \bibitem{} Kuijken K. \& Gilmore G., 1989, MNRAS 239, 605 
   
   \bibitem{} Lacey C.G. \& Fall M., 1985, ApJ 290, 154 
   
   \bibitem{} Larson R.B., 1974, MNRAS 166, 585 
   
   \bibitem{} Larson R.B., 1976, MNRAS 176, 31 
   
   \bibitem{} Lyne A., Manchester R. \& Taylor J., 1985, MNRAS 213, 613
    
   \bibitem{} Majewski S.K., 1993, ARA\&A 31, 575 
   
   \bibitem{} Malinie G., Hartmann D.H., Clayton D.D. \& Mathews G.J., 1993, ApJ 413, 633
   
   \bibitem{} Matteucci F. \& Francois P., 1989, MNRAS 239, 885 
   
   \bibitem{} Matteucci F. \& Greggio, L. 1986, A\&A 154, 279 
   
   \bibitem{} Mo H.J., Mao S. \& White S.D.M., 1998, preprint (astro-ph 9807341) 
   
   \bibitem{} Nissen P.E. \& Schuster W.J., 1997, A\&a 326, 751
   
   \bibitem{} Nordstrom B., Olsen E.H. \& Andersen J., 1997, in ASP Conf. Series, 112, The History of the Milky Way, ed. A. Burkert, 
   D.H. Hartmann \& S.R. Majewski, p.\ 144 
     
   \bibitem{} Norris J.E., 1987, ApJ 314, L39 
   
   \bibitem{} Ojha D.K., Bienayme O., Robin A.C. \& Mohan V., 1994a, A\&A 284, 810
   
   \bibitem{} Ojha D.K., Bienayme O., Robin A.C. \& Mohan V., 1994b, A\&A 290, 771
   
   \bibitem{} Ojha D.K., Bienayme O., Robin A.C., Creze M. \& Mohan V., 1996, A\&A 311, 456 
   
   \bibitem{} Olsen E.H., 1993, A\&AS 102, 89
   
   \bibitem{} Pagel B.E.J., 1997, Nucleosynthesis and Galactic Chemical Evolution 
   (Cambridge: Cambridge Univ. Press)
   
   \bibitem{} Pagel B.E.J. \& Patchett B.E., 1975, MNRAS 172, 13 
   
   \bibitem{} Pagel B.E.J., 1989, in Evolutionary Phenomena in Galaxies, ed. J. Beckman \& B.E.J.  
   Pagel ( Cambridge: Cambridge Univ. Press), p.\ 201 
   
   \bibitem{} Pilyugin L.S. \& Edmunds M.G., 1996a, A\&A 313, 792 
   
   \bibitem{} Pilyugin L.S. \& Edmunds M.G., 1996b, A\&A 313, 783 
   
   \bibitem{} Prantzos N. \& Aubert O., 1995, A\&A, 302, 69
   
   \bibitem{} Prantzos N. \& Silk J., 1998, ApJ 507, 229
   
   \bibitem{} Primack J.R., Bullock J.S., Somerville R.s. \& MacMinn D., 1999, preprint (astro-ph 9812141)
   
   \bibitem{} Quinn P.J., Hernquist L. \& Fullager D.P., 1993, ApJ 403, 74
   
   \bibitem{} Rana N.C., 1991, ARA\&A 29, 129 
   
   \bibitem{} Reid N. \& Majewski S.R., 1993, ApJ 409, 635 
   
   \bibitem{} Robin A.C., Haywood M., Creze M., Ojha D.K. \& Bienayme O., 1996, A\&A 305,125 
   
   \bibitem{} Rose J.A., 1985, AJ 90,787 
   
   \bibitem{} Rocha-Pinto H.J. \& Maciel W.J., 1996, MNRAS 279, 447
   
   \bibitem{} Sackett P.D., 1997, ApJ 483, 103 
   
   \bibitem{} Samland M., Hensler G. \& Theis C.H., 1997, ApJ 476, 544 
   
   \bibitem{} Sandage A., 1990, J.R.Astron.Soc.Con. 84, 70 
   
   \bibitem{} Sandage A., 1987, AJ 93, 610 
   
   \bibitem{} Scoville N. \& Sanders D., 1987, in "Interstellar Processes", Eds. H. Thronson \& D. Hollenbach (Kluwer) P.\ 21

   \bibitem{} Shaver P.A., McGee R.X., Newton L.M., Danks A.C. \& Pottasch S.R., 1983,  
   MNRAS 204, 53 
   
   \bibitem{} Smartt S.J. \& Rolleston W.R.J., 1997, ApJ 481, L47 
   
   \bibitem{} Sommer-Larsen J., 1991, MNRAS 249, 368

   \bibitem{} Somerville R.S. \& Primack J.R., 1998, MNRAS in press (astro-ph 9802268) 
   
   \bibitem{} Thon R., \& Meusinger H., 1998, A\&A 338, 413 
   
   \bibitem{} Timmes, T.X., Woosley, S.E. \& Weaver, T.A., 1995, ApJS 98, 617
   
   \bibitem{} Tinsley B., 1980, Fundam. Cosmic Phys. 5, 287 
   
   \bibitem{} Tosi M., 1996, in ASP Conf. Proc. 98, From Stars to Galaxies, ed. C. Leitherer, U.  
   Fritze-von Alvensleben \& J. Huchra (San Francisco: ASP), p.\ 299 
   
   \bibitem{} Twarog B.A., 1980, ApJ 242, 242 
   
   \bibitem{} van den Hoek L.B. \& Groenewegen M.A.T., 1997, A\&AS 123, 305 
   
   \bibitem{} Vilchez J.M. \& Esteban C., 1996, MNRAS 280, 720 
   
   \bibitem{} Wang B. \& Silk J., 1994, ApJ 427, 759 
   
   \bibitem{} Wechsler R.H., Gross M.A.K., Primack J.R., et al 1998, preprint (astro-ph 9712141)
   
   \bibitem{} Weideman V., 1984, A\&A 134, L1 
  
   \bibitem{} White S.D.M. \& Frenck C.S>, 1991, ApJ 379, 52
   
   \bibitem{} White S.D.M. \& Rees M.J., 1978, MNRAS 183, 341
   
   \bibitem{} Woosley S.E. \& Weaver T.A., 1995, ApJS 101, 181 
   
   \bibitem{} Woosley S.E., 1997, ApJ 476, 801 

\end{thebibliography}
\end{document}